\documentclass[11pt]{article}
\usepackage{amssymb,cite}
\makeatletter
\@addtoreset{equation}{section}
\makeatother

\def\ben{\begin{equation}}
\def\een{\end{equation}}

  \let\n=\nu

\let\C=\Chi

\def\nn{\nonumber} \def\bd{\begin{document}} \def\ed{\end{document}}
\def\ds{\documentstyle} \let\fr=\frac \let\bl=\bigl \let\br=\bigr
\let\Br=\Bigr \let\Bl=\Bigl
\let\bm=\bibitem
\let\na=\nabla
\let\pa=\partial \let\ov=\overline
\newcommand{\be}{\begin{equation}}
\newcommand{\ee}{\end{equation}}
\def\ba{\begin{array}}
\def\ea{\end{array}}
\def\ft#1#2{{\textstyle{{\scriptstyle #1}\over {\scriptstyle #2}}}}
\def\fft#1#2{{#1 \over #2}}
\def\del{\partial}
\def\vp{\varphi}
\def\sst#1{{\scriptscriptstyle #1}}
\def\oneone{\rlap 1\mkern4mu{\rm l}}
\def\td{\tilde}
\def\wtd{\widetilde}
\def\ie{\rm i.e.\ }
\def\dalemb#1#2{{\vbox{\hrule height .#2pt
        \hbox{\vrule width.#2pt height#1pt \kern#1pt
                \vrule width.#2pt}
        \hrule height.#2pt}}}
\def\square{\mathord{\dalemb{6.8}{7}\hbox{\hskip1pt}}}
\newcommand{\ho}[1]{$\, ^{#1}$}
\newcommand{\hoch}[1]{$\, ^{#1}$}
\newcommand{\bea}{\begin{eqnarray}}
\newcommand{\eea}{\end{eqnarray}}
\newcommand{\ra}{\rightarrow}
\newcommand{\lra}{\longrightarrow}
\newcommand{\Lra}{\Leftrightarrow}
\newcommand{\ap}{\alpha^\prime}
\newcommand{\bp}{\tilde \beta^\prime}
\newcommand{\tr}{{\rm tr} }
\newcommand{\Tr}{{\rm Tr} }
\def\0{{\sst{(0)}}}
\def\1{{\sst{(1)}}}
\def\2{{\sst{(2)}}}
\def\3{{\sst{(3)}}}
\def\4{{\sst{(4)}}}
\def\5{{\sst{(5)}}}
\def\6{{\sst{(6)}}}
\def\7{{\sst{(7)}}}
\def\8{{\sst{(8)}}}
\def\n{{\sst{(n)}}}
\def\cA{{{\cal A}}}
\def\cB{{{\cal B}}}
\def\cF{{{\cal F}}}
\def\cH{{{\cal H}}}
\def\tV{\widetilde V}
\def\tW{\widetilde W}
\def\tH{\widetilde H}
\def\tE{\widetilde E}
\def\tF{\widetilde F}
\def\tA{\widetilde A}
\def\im{{{\rm i}}}
\def\tY{{{\wtd Y}}}
\def\ep{{\epsilon}}
\def\vep{{\varepsilon}}
\def\R{\rlap{\rm I}\mkern3mu{\rm R}}
\def\bD{{{\bar D}}}
\def\R{\rlap{\rm I}\mkern3mu{\rm R}}
\def\bD{{{\bar D}}}
\def\R{{{\mathbb R}}}
\def\C{{{\mathbb C}}}
\def\H{{{\mathbb H}}}
\def\CP{{{\mathbb C}{\mathbb P}}}
\def\RP{{{\mathbb R}{\mathbb P}}}
\def\Z{{{\mathbb Z}}}
\def\bA{{{\mathbb A}}}
\def\bB{{{\mathbb B}}}
\def\bC{{{\mathbb C}}}
\def\bD{{{\mathbb D}}}
\def\bE{{{\mathbb E}}}
\def\bZ{{{\mathbb Z}}}
\def\Re{{{\mathfrak{Re}}}}
\def\Im{{{\mathfrak{Im}}}}
\def\cosec{{\,\hbox{cosec}\,}}
\def\Gm{{\Gamma_{\!\! -}}}
\def\Gp{{\Gamma_{\!\! +}}}
\def\stan{{standard }}
\def\nonstan{{supernumerary }}
\def\FF2{{ {}_{\sst 2}F_{\sst 1} }}
\def\FFF{{ {}_{\sst 3}F_{\sst 2} }}
\def\const{\rm constant}

\newcommand{\auth}{Muraari Vasudevan$^*$ and Kory A. Stevens$^\dagger$}

\begin{document}

\begin{flushright}

Alberta Thy 10-05\\
PACS numbers: 04.50.+h, 98.80.Cq\hfill\\
July\  2005
\end{flushright}

\vspace{10pt}

\begin{center}
{\large {\bf Integrability of Particle Motion and Scalar Field
Propagation in Kerr-(Anti) de Sitter Black Hole Spacetimes in All
Dimensions
            }}

\vspace{20pt}
\auth

\vspace{10pt}

{\it $^*$ Theoretical Physics Institute, University of Alberta,\\
Edmonton, Alberta  T6G 2J1, Canada}

{\it
    {\rm E-mail: \texttt{mvasudev@phys.ualberta.ca}}
  }

{\it $^\dagger$ Department of Physics and Astronomy, University of British Columbia,\\
Vancouver, British Columbia  V6T 1Z1, Canada}

{\it
   {\rm E-mail: \texttt{kstevens@phas.ubc.ca}}
  }


\vspace{40pt}

\underline{ABSTRACT}
\end{center}

We study the Hamilton-Jacobi and massive Klein-Gordon equations in
the general Kerr-(Anti) de Sitter black hole background in all
dimensions. Complete separation of both equations is carried out
in cases when there are two sets of equal black hole rotation
parameters. We analyze explicitly the symmetry properties of these
backgrounds that allow for this Liouville integrability and
construct a nontrivial irreducible Killing tensor associated with
the enlarged symmetry group which permits separation. We also
derive first-order equations of motion for particles in these
backgrounds and examine some of their properties. This work
greatly generalizes previously known results for both the
Myers-Perry metrics, and the Kerr-(Anti) de Sitter metrics in
higher dimensions.

\pagebreak

\section{Introduction}

A number of recent developments in high energy physics have
generated great interest in vacuum solutions of Einstein equations
describing higher dimensional black holes, and the properties of
these spacetimes. Models of spacetimes with large extra dimensions
have been proposed to deal with several questions arising in
modern particle phenomenology (e.g. the hierarchy problem)
\cite{NDD, ANDD, RS}. Higher dimensional black hole solutions
arise naturally in such models. These models are also of interest
in the context of mini-black hole production in high energy
particle colliders, which would provide a window into
non-perturbative gravitational physics \cite{Cav, Kan}.

Superstring and M-theory also naturally give rise to higher
dimensional black holes in their 10 or 11 dimensional ambient
spacetimes. P-branes present in these theories can also support
black holes, thereby making black hole solutions in an
intermediate number of dimensions physically interesting as well.
Solitonic objects in superstring theory frequently find a natural
description in terms of higher dimensional black holes. They
provide important keys to understanding strongly coupled
non-perturbative phenomena which cannot be ignored at the
Planck/string scale \cite{DV, CT}.

The Kerr metric describes astrophysically relevant black hole
spacetimes, to a very good approximation\cite{Kerr}. One
generalization of the Kerr metric to higher dimensions is given by
the Myers-Perry construction \cite{MyersPerry}. With interest now
in a nonzero cosmological constant, it is worth studying
spacetimes describing rotating black holes with a cosmological
constant. Another motivation for including a cosmological constant
is driven by the AdS/CFT correspondence. The study of black holes
in an Anti-de Sitter background could give rise to interesting
descriptions in terms of the conformal field theory on the
boundary leading to better understanding of the correspondence
\cite{Mal, Wit}. The general Kerr-de Sitter metrics describing
rotating black holes in the presence of a cosmological constant
have been constructed explicitly in \cite{GPLP, GPLP2}.

There is also a very strong need to understand the structure of
geodesics in the background of black holes in Anti-de Sitter
backgrounds in the context of string theory and the AdS/CFT
correspondence. This is due to the recent work in exploring black
hole singularity structure using geodesics and correlators on the
dual CFT on the boundary \cite{Shen1, Shen2, Shen3, Shen4, Shen5,
Shen6}.

In this paper we study the separability of the Hamilton-Jacobi
equation in these spacetimes, which can be used to describe the
motion of classical massive and massless particles (including
photons). We also investigate the separability of the Klein-Gordon
equation describing a massive scalar field propagating in this
background. We explicitly perform the separation in the case where
there are only two sets of equal rotation parameters describing
the black hole. We use this explicit separation to obtain
first-order equations of motion for both massive and massless
particles in these backgrounds. The equations are obtained in a
form  that could be used for numerical study, and also in the
study of black hole singularity structure using geodesic probes
and the AdS/CFT correspondence.

We also study the Klein-Gordon equation describing the propagation
of a massive scalar field in this spacetime. Separation is again
explicitly shown for the case of two sets of equal black hole
rotation parameters. We construct the separation of both equations
explicitly in these cases.

This paper greatly generalizes the results of \cite{frolov1 ,
frolov2} for the Myers-Perry metric in five dimensions,
\cite{VSP1} which separates the equations in the case of equal
rotation parameters in the odd dimensional Kerr-(A)dS spacetimes,
and \cite{VSP2} which separates the equations in the case of two
independent sets of rotation parameters in the Myers-Perry metrics
in all dimensions, as well as some related results in five
dimensional black hole spacetimes in \cite{MV,KL}.

Separation is possible for both equations in this case due to the
existence of a second-order non-trivial irreducible Killing
tensor. This is a generalization of the Killing tensor in the Kerr
black hole spacetime in four dimensions constructed in
\cite{Carter} which was subsequently described by Chandrasekhar as
the ``miraculous property of the Kerr metric".  A similar
construction for the Myers-Perry metrics in higher dimensions has
also been done\cite{frolov1, VSP2}. The Killing tensor provides an
additional integral of motion necessary for complete
integrability. We also construct Killing vectors, which exist due
to the additional symmetry, and which permit the separation of
these equations.

\section{Construction and Overview of the Kerr-de Sitter Metrics}

One of the most useful properties of the Kerr metric is that it
can be written in the Kerr-Schild \cite{KerrSchild} form, where
the metric $g_{\mu \nu}$ is given exactly by its linear
approximation around the flat metric $\eta _{\mu \nu} $ as
follows: \ben ds^2 =g_{\mu \nu} dx^\mu dx^\nu  = \eta_{\mu \nu}
dx^\mu dx^\nu
 + {2 M \over U} \, (k_\mu dx^\mu)^2\,,\label{ks1}
\een where $k_\mu $ is null and geodesic with respect to both the
full metric
 $g_{\mu \nu}$ and the flat metric $\eta _{\mu \nu}$.

The Kerr-de Sitter metrics in all dimensions were obtained in
\cite{GPLP} by using the de Sitter metric instead of the flat
background $\eta _{\mu \nu}$, with coordinates chosen
appropriately to allow for the incorporation of the Kerr metric
via the null geodesic vectors $k_{\mu}$. We quickly review the
construction here.

In $D$-dimensional spacetime, we introduce $n=[D/2]$ coordinates $\mu_i$, where
$[i]$ denotes the integer part of $i$, subject to the constraint
\be \sum_{i=1}^n \mu_i^2 =1\,, \label{constraint} \ee
together with $N=[(D-1)/2]$ azimuthal angular coordinates
$\phi_i$, the radial coordinate $r$, and the time coordinate $t$.
When the total spacetime dimension $D$ is odd, $D=2n+1=2N+1$,
there are $n$ azimuthal coordinates $\phi_i$, each with period
$2\pi$.  If $D$ is even, $D=2n=2N+2$, there are only $N=n-1$
azimuthal coordinates $\phi_i$. Define $\epsilon$ to be 1 for even
$D$, and 0 for odd $D$.

The Kerr-de Sitter metric $ds^2$ in $D$ dimensional spacetime
satisfies the Einstein equation with cosmological constant
$\lambda$:
\be R_{\mu\nu} = (D-1)\, \lambda\, g_{\mu\nu}\,. \ee
Define functions $W$ and $F$ as follows:
\be W \equiv \sum_{i=1}^n \fft{\mu_i^2}{1+\lambda\,
a_i^2}\,,\qquad 
F\equiv \fft{r^2}{1-\lambda\, r^2}\, \,
  \sum_{i=1}^n \fft{\mu_i^2}{r^2+a_i^2}\,.\label{WFdef}
\ee
In $D$ dimensions, the Kerr-de Sitter metrics are given by
\be ds^2 = d\bar s^2 + \fft{2M}{U}\, (k_\mu\,
dx^\mu)^2\,,\label{kds} \ee
where the de Sitter metric $d\bar s^2$, the null vector $k_\mu$,
and the function $U$ are now given by
\bea d\bar s^2 &=& - W\,(1-\lambda \, r^2)\, dt^2 + F\, dr^2 +
\sum_{i=1}^n \fft{r^2 + a_i^2}{1+\lambda\, a_i^2} \,\, d\mu_i^2 +
\sum_{i=1}^{n-\epsilon} \fft{r^2 + a_i^2}{1+\lambda\, a_i^2}
\, \, \mu_i^2\, d\phi_i^2 \nn \\
&& \qquad +
 \fft{\lambda}{W\, (1-\lambda\, r^2)}\,
\Big( \sum_{i=1}^n \fft{(r^2 + a_i^2)\, \mu_i\,
d\mu_i}{1+\lambda\, a_i^2}
   \Big)^2\,,\label{case}\\
k_\mu\, dx^\mu &=& W\, dt +  F\, dr - \sum_{i=1}^{n-\epsilon}
\fft{a_i\, \mu_i^2}{1+\lambda\, a_i^2} \,  d\phi_i \,,
\label{case2}\\
U &=& r^{\epsilon}\, \sum_{i=1}^n \fft{\mu_i^2}{r^2 + a_i^2}\,
\prod_{j=1}^{n-\epsilon} (r^2 + a_j^2)\,.\label{Udef} \eea
In the even-dimensional case, where there is no azimuthal
coordinate $\phi_n$, there is also no associated rotation
parameter; i.e., $a_n=0$. Note that the null vector corresponding
to the null one-form is
\be k^\mu\, \del_\mu = -\fft{1}{1-\lambda\, r^2}\, \fft{\del}{\del
t}
    + \fft{\del}{\del r} - \sum_{i=1}^{n-\epsilon}\, \fft{a_i}{r^2+ a_i^2}\,
\fft{\del}{\del \phi_i}\,. \ee
This is easily obtained by using the background metric to raise
and lower indices rather than the full metric, since $k$ is null
with respect to both metrics.

For the purposes of analyzing the equations of motion and the
Klein-Gordon equation, it is very convenient to work with the
metric expressed in Boyer-Lindquist coordinates. In these
coordinates there are no cross terms involving the differential
$dr$. In both even and odd dimensions, the Boyer-Lindquist form is
obtained by means of the following coordinate transformation:
\be dt=d\tau + \fft{2M\, dr}{(1 - \lambda\, r^2)(V-2M)}\,,\qquad
d\phi_i = d\varphi_i - \lambda\,a_i\,d\tau + \fft{2M\,a_i\,
dr}{(r^2 + a_i^2)(V-2M)}\,.\label{coordtrans} \ee
In Boyer-Lindquist coordinates in $D$ dimensions, the Kerr-de
Sitter metrics are given by

\bea ds^2 &=& - W\, (1 - \lambda r^2)\, d\tau^2 + \fft{U\,
dr^2}{V-2M} + \fft{2M}{U}\Bigl(d\tau - \sum_{i=1}^{n-\epsilon}
\fft{a_i\, \mu_i^2\, d\varphi_i}{
1 + \lambda\, a_i^2}\Bigr)^2 \nn\\
&&+ \sum_{i=1}^n \fft{r^2 + a_i^2}{1 + \lambda\, a_i^2} \,
d\mu_i^2 + \sum_{i=1}^{n-\epsilon} \fft{r^2 + a_i^2}{1 + \lambda\,
a_i^2}\,
\mu_i^2\, (d\varphi_i-\lambda\, a_i\, d\tau)^2 \nn\\
&&+ \fft{\lambda}{W\, (1-\lambda r^2)} \Big( \sum_{i=1}^n
\fft{(r^2 + a_i^2)\mu_i\, d\mu_i}{ 1 + \lambda\, a_i^2}\Big)^2
\,,\label{bl} \eea
where $V$ is defined here by
\be V\equiv  r^{\epsilon -2}(1-\lambda
r^2)\prod_{i=1}^{n-\epsilon} (r^2 + a_i^2) =
\fft{U}{F}\,,\label{Vdef} \ee
Note that obviously $a_n=0$ in the even dimensional case, as there
is no rotation associated with the last direction.

\section{Inverting the Kerr-(A)dS metric in all dimensions}

We briefly review the process of inversion of the metric using the
Kerr-Schild formalism. More extensive details of this type of
procedure can be found in \cite{VSP1, VSP2}. This section will
also help establish some useful notation and conventions for the
rest of the paper. Note that the metric is block diagonal in the
$(\mu _i)$ and the $(r,\tau ,\varphi _i)$ sectors and so can be
inverted separately.

To deal with the $(r,\tau ,\varphi _i)$ sector, the most efficient
method is to use the Kerr-Schild construction of the metric. From
(\ref{ks1}) and using the fact that $k$ is null, we can write \ben
g^{\mu \nu}=\eta ^{\mu \nu} -\frac{2M}{U}k^{\mu}k^{\nu} \,, \een
where $\eta$ here is the de Sitter metric rather than the flat
metric, and we raise and lower indices with $\eta$. Since the null
vector $k$ has no components in the $\mu _i$ sector, we can regard
the above equation as holding true in the $(r,\tau ,\varphi _i)$
sector with $k$ null here as well. Then we can explicitly perform
the coordinate transformation (\ref{coordtrans}) (or rather its
inverse) on the raised metric to obtain the components of $g^{\mu
\nu}$ in Boyer-Lindquist coordinates in the $(r,\tau,\varphi _i)$
sector.

We get the following components for the $(r,\tau ,\varphi _i)$
sector of $g^{\mu \nu}$:
\begin{eqnarray}
g^{\tau r}&=&g^{\varphi _i r}=0 \,, \nonumber \\
g^{rr}&=&\frac{V-2M}{U}\,, \nonumber \\
g^{\tau \tau}&=&Q-\frac{4M^2}{U(1-\lambda r^2)^2(V-2M)}\,, \nonumber \\
g^{\tau \varphi _i}&=&\lambda a_i Q -\frac{4M^2a_i(1+\lambda a_i
^2)}{U(1-\lambda r^2)^2 (V-2M)(r^2+a_i ^2)}
-\frac{2M}{U}\frac{a_i}{(1-\lambda
r^2)(r^2+a_i ^2)}\,, \nonumber \\
g^{\varphi _i \varphi_j}&=& \frac{(1+\lambda a_i ^2)}{(r^2+a_i
^2)\mu _i ^2}
\delta ^{ij}+\lambda ^2 a_i a_j Q +\frac{Q^{ij}}{U}  \nonumber \\
&&+\frac{4M^2 a_i a_j (1+\lambda a_i^2) (1+\lambda a_j
^2)}{U(1-\lambda r^2)^2(V-2M)(r^2+a_i ^2)(r^2 +a_j ^2)}  \,,
\label{rtfinv}
\end{eqnarray}
where $Q$ and $Q^{ij}$ are defined to be \ben
Q=-\frac{1}{W(1-\lambda r^2)}-\frac{2M}{U}\frac{1}{(1-\lambda
r^2)^2} \,, \label{qlam} \een

\begin{eqnarray}
Q^{ij}=\frac{-4M^2\lambda a_i a_j[(1+\lambda a_j ^2)(r^2+a_i
^2)+(1+\lambda a_i^2)(r^2+a_j ^2)]}{(1-\lambda r^2)^2
(V-2M)(r^2+a_i^2)(r^2+a_j ^2)}-2M\frac{a_ia_j}{(r^2+a_i
^2)(r^2+a_j ^2)}
\nonumber \\
-\frac{2M\lambda a_i a_j}{(1-\lambda r ^2)} \left[ \frac{1}{(r^2 +
a^2 _i)}+\frac{1}{(r^2 + a^2 _j )}\right] - \frac{4M^2 a_i
a_j[(1+\lambda a_i ^2)+(1+\lambda a_j^2)]}{(1-\lambda r^2)^2
(V-2M)(r^2+a_i^2)(r^2+a_j^2)} \,.
 \label{qij}
\end{eqnarray}

These results were compared to previously known ones in the case
of $\lambda =0$ and showed agreement \cite{frolov1}. Also, we used
the GRTensor package for Maple explicitly to check that this is
the correct inverse metric \cite{Map}.

Note that the functions $W$ and $U$ both depend explicitly on the
$\mu_i$'s. Unless the $(r,\tau,\varphi_i)$ sector can be decoupled
from the $\mu$ sector, complete separation is unlikely. If
however, all the $a_i=a$ for some non-zero value $a$, then $W$ and
$U$ are no longer $\mu$ dependent (taking the constraint into
account) and separation seems likely. Note, however, that in this
case we cannot deal with even dimensional spacetimes, since
$a_n=0$ is different from the other $a_i=a$. The analysis in this
case has been done in detail in \cite{VSP1}.

We will actually work with a much more general case, in which
separation works in both even and odd dimensional spacetimes. We
consider the situation in which the set of rotation parameters
$a_i$ take on at most only two distinct values $a$ and $b$ ($a=b$
can be obtained as a special case). In even dimensions at least
one of these values must be zero, since $a_n=0$. As such in even
dimensions we take $b=0$ and $a$ to be some (possibly different)
value. In the odd dimensional case, there are no restrictions on
the values of $a$ and $b$. We adopt the convention
\begin{eqnarray}
a_i = a \qquad \textrm{for} \quad i=1,...,m \quad, \qquad a_{j+m} = b
\qquad \textrm{for} \quad j=1,...,p \, ,
\end{eqnarray}
where $m+p=N+\epsilon=n$.

Since the $\mu_i$'s are constrained by (\ref{constraint}), we need
to use suitable independent coordinates instead. We use the
following decomposition of the $\mu_i$:
\begin{equation}
\mu_i = \lambda _i \sin \theta \quad \textrm{for} \quad i=1,...,m
\quad, \qquad \mu_{j+m} = \nu _j \cos\theta \quad \textrm{for}
\quad j=1,...,p \, , \label{mudef}
\end{equation}
where the $\lambda_i$ and $\nu_j$ have to satisfy the constraints
\begin{equation}
\sum_{i=1}^{m} \lambda _i ^2 =1 \qquad , \qquad \sum_{j=1}^{p} \nu
_ j ^2 =1 \, .
\end{equation}

Since these constraints describe unit $(m-1)$ and $(p-1)$
dimensional spheres in the $\lambda$ and $\nu$ spaces
respectively, the natural choice is to use two sets of spherical
polar coordinates. We write
\begin{eqnarray}
\lambda_i&=& \left( \prod _{k=1}^{m-i} \sin\alpha _k\right) \cos \alpha _{m-i+1} \,, \nonumber \\
\nu_j&=& \left( \prod _{k=1}^{p-j} \sin\beta _k\right) \cos \beta
_{p-j+1} \,, \label{lnsphere}
\end{eqnarray}
with the understanding that the products are one when $i=m$ or
$j=p$ respectively, and that $\alpha _m=0$ and $\beta _p =0$.

The $\mu$ sector metric can then be written as
\begin{eqnarray}
ds_{\mu}^2&=&\frac{\rho ^2}{\Delta _\theta} d\theta ^2 + \frac{r^2+a^2}{\Sigma_a} \sin ^2 \theta \sum_{i=1}^{m-1} \left( \prod _{k=1} ^{i-1} \sin ^2 \alpha _k \right) d\alpha _i ^2 \nonumber \\
&&\qquad +\frac{r^2+b^2}{\Sigma_b} \cos ^2 \theta \sum_{j=1}^{p-1}
\left( \prod _{k=1} ^{j-1} \sin ^2 \beta _k \right) d\beta _j ^2
\, , \label{musector}
\end{eqnarray}
again with the understanding that the products are one when $i=1$
or $j=1$, and we use the definitions
\begin{eqnarray}
\rho ^2 &=& r^2 + a^2 \cos ^2 \theta + b^2 \sin ^2 \theta \, ,
\nonumber \\
\Delta_\theta&=&1+\lambda a^2\cos ^2\theta + \lambda b^2 \sin
^2\theta \, , \nonumber \\
\Sigma_a &=& 1+\lambda a^2 \, , \nonumber \\
\Sigma_b &=& 1+\lambda b^2 \, , \nonumber \\
Z&=&r^\epsilon (r^2+a^2)^{m-1} (r^2+b^2)^{p-1-\epsilon} \, .
\label{newfns}
\end{eqnarray}
This diagonal metric can be easily inverted to give
\begin{eqnarray}
g^{\theta \theta} &=& \frac{\Delta_\theta}{\rho ^2} \, , \nonumber \\
g^{\alpha _i \alpha _j} &=& \frac{\Sigma_a}{(r^2+a^2)\sin ^2
\theta} \frac{1}{\left(\prod
_{k=1} ^{i-1} \sin ^2 \alpha _k\right)}\delta _{ij} \, , \qquad i,j=1,...,m \, , \nonumber \\
g^{\beta _i \beta _j} &=& \frac{\Sigma_b}{(r^2+b^2)\cos ^2 \theta}
\frac{1}{\left(\prod _{k=1} ^{i-1} \sin ^2 \beta _k\right)}\delta
_{ij} \, , \qquad i,j =1,...,p \,. \label{muinv}
\end{eqnarray}

For the case of two sets of rotation parameters that we consider
here, the following expressions will be extremely useful:
\begin{eqnarray}
U&=&\rho ^2 Z\, , \nonumber \\
W&=&\frac{\Delta_\theta}{\Sigma _a \Sigma _b} \, .
\end{eqnarray}
We note that both $V$ and $Z$ are functions of $r$ only.

The following identity, which can be easily verified, will be
crucial in the following:
\begin{eqnarray}
Q=\frac{\Sigma _a \Sigma _b}{\rho ^2 \lambda \Delta _\theta}
-\frac{\Sigma _a \Sigma _b}{\rho ^2 \lambda (1-\lambda
r^2)}-\frac{2M}{\rho ^2 Z (1-\lambda r^2)^2 (V-2M)} \, .
\label{Qid}
\end{eqnarray}

\section{The Hamilton-Jacobi Equation and Separation}
The Hamilton-Jacobi equation in a curved background is given by
\be -\frac{\partial S}{\partial l} = H = \frac{1}{2} g^{\mu \nu }
\frac{\partial S}{\partial x^{\mu}} \frac{\partial S}{\partial
x^{\nu}} \,, \label{HJ} \ee where $S$ is the action associated
with the particle and $l$ is some affine parameter along the
worldline of the particle. Note that this treatment also
accommodates the case of massless particles, where the trajectory
cannot be parameterized by proper time.

We can attempt a separation of coordinates as follows. Let
\begin{equation}
S=\frac{1}{2}m^2 l -E\tau + \sum_{i=1} ^m \Phi_i \varphi _i
+\sum_{i=1}^{p} \Psi_i \varphi _{m+i} +S_r (r) + S_{\theta}(\theta)
+\sum _{i=1} ^{m-1} S_{\alpha _i} (\alpha _i) +\sum _{i=1} ^{p-1}
S_{\beta _i} (\beta _i)\,. \label{ansatz}
\end{equation}
$\tau$ and $\varphi _i$ are cyclic coordinates, so their conjugate momenta are
conserved. The conserved quantity associated with time translation is the energy
$E$, and the conserved quantity associated with rotation in each $\varphi _i$ is the corresponding angular momentum $\Phi_i$ or $\Psi_j$. We also adopt the convention that $\Psi_p=0$ in an even
number of spacetime dimensions.

Using (\ref{rtfinv}), (\ref{newfns}), (\ref{muinv}), and
(\ref{ansatz}) we write the Hamilton-Jacobi equation (\ref{HJ}) as
\begin{eqnarray}
-m^2 = && \left[\frac{\Sigma_a \Sigma _b}{\lambda\rho ^2 \Delta_\theta}-\frac{\Sigma_a\Sigma_b}{\rho^2\lambda(1-\lambda r^2)}-\frac{2M}{\rho ^2 Z(1-\lambda r^2)}-\frac{4M^2}{\rho^2Z(1-\lambda r^2)^2}\right]E^2 \nonumber \\
&+&2\left[\frac{a\Sigma_a \Sigma _b}{\rho ^2 \Delta_\theta}-\frac{a\Sigma_a\Sigma_b}{\rho^2(1-\lambda r^2)}-\frac{2M\lambda a}{\rho ^2 Z(1-\lambda r^2)^2}-\frac{4M^2a\Sigma_a}{\rho^2Z(1-\lambda r^2)^2(V-2M)(r^2+a^2)}\right. \nonumber \\
&-&\left.\frac{2Ma}{\rho^2Z(1-\lambda r^2) (r^2+a^2)} \right]\sum_{i=1}^m (-E){\Phi_i} \nonumber \\
&+&2\left[\frac{b\Sigma_a \Sigma _b}{\rho ^2 \Delta_\theta}-\frac{b\Sigma_a\Sigma_b}{\rho^2(1-\lambda r^2)}-\frac{2M\lambda b}{\rho ^2 Z(1-\lambda r^2)^2}-\frac{4M^2b\Sigma_b}{\rho^2Z(1-\lambda r^2)^2(V-2M)(r^2+b^2)}\right. \nonumber \\
&-&\left.\frac{2Mb}{\rho^2Z(1-\lambda r^2) (r^2+b^2)} \right]\sum_{j=1}^p (-E){\Psi_j} \nonumber \\
&+& \frac{\Sigma_a}{(r^2+a^2)\sin^2\theta} \sum_{i=1}^m \frac{\Phi
_ i ^2}{\lambda _i^2}+\frac{\Sigma_b}{(r^2+b^2)\cos ^2 \theta}
\sum_{i=1}^p
\frac{\Psi _ i^2}{\nu_i ^2} +\frac{\Delta_\theta}{\rho ^2}\left[\frac{dS_\theta(\theta)}{d\theta}\right]^2\nonumber \\
&+&\frac{V-2M}{\rho ^2 Z} \left[\frac{dS_r(r)}{dr}\right]^2+ \sum_{i=1} ^{m-1} \frac{\Sigma_a}{(r^2+a^2)\sin ^2 \theta \prod _{k=1} ^{i-1} \sin ^2 \alpha _k } \left( \frac{dS_{\alpha _i}}{d\alpha _i}\right) ^2 \nonumber \\
&+& \sum_{i=1} ^{p-1} \frac{\Sigma _b}{(r^2+b^2)\cos ^2 \theta
\prod _{k=1} ^{i-1} \sin ^2 \beta _k } \left( \frac{dS_{\beta
_i}}{d\beta _i}\right) ^2 \nonumber \\
&+& \sum_{i=1}^m\sum_{j=1}^m
\left[\lambda^2a^2\left(\frac{\Sigma_a \Sigma _b}{\rho ^2
\Delta_\theta}-\frac{\Sigma_a\Sigma_b}{\rho^2\lambda(1-\lambda
r^2)}-\frac{2M}{\rho ^2 Z(1-\lambda r^2)}\right) \right. \nonumber \\
&+& \left. \frac{4M^2a^2\Sigma_a^2}{\rho^2
Z(V-2M)(r^2+a^2)^2}+\frac{Q^{ij}}{\rho^2Z}\right]\Phi_i \Phi_j \nonumber \\
&+& \sum_{i=1}^p\sum_{j=1}^p
\left[\lambda^2b^2\left(\frac{\Sigma_a \Sigma _b}{\rho ^2
\Delta_\theta}-\frac{\Sigma_a\Sigma_b}{\rho^2\lambda(1-\lambda
r^2)}-\frac{2M}{\rho ^2 Z(1-\lambda r^2)}\right) \right. \nonumber \\
&+& \left. \frac{4M^2b^2\Sigma_a^2}{\rho^2
Z(V-2M)(r^2+b^2)^2}+\frac{Q^{(i+m)(j+m)}}{\rho^2Z}\right]\Psi_i \Psi_j \nonumber \\
&+& 2\sum_{i=1}^m\sum_{j=1}^p
\left[\lambda^2ab\left(\frac{\Sigma_a \Sigma _b}{\rho ^2
\Delta_\theta}-\frac{\Sigma_a\Sigma_b}{\rho^2\lambda(1-\lambda
r^2)}-\frac{2M}{\rho ^2 Z(1-\lambda r^2)}\right) \right. \nonumber \\
&+& \left. \frac{4M^2ab\Sigma_a\Sigma_b}{\rho^2
Z(V-2M)(r^2+a^2)(r^2+b^2)}+\frac{Q^{i(j+m)}}{\rho^2Z}\right]\Phi_i\Psi_j\,.
\end{eqnarray}

Note that here the $\lambda_i$ and $\nu_j$ are not coordinates,
but simply quantities defined by (\ref{lnsphere}). We continue to
use the convention defined for products of $\sin ^2 \alpha _i$ and
$\sin ^2 \beta _j$ defined earlier. Separate the $\alpha_i$ and
$\beta _j$ coordinates from the Hamilton-Jacobi equation via
\begin{eqnarray}
J_1 ^2&=&\sum_{i=1}^m \left[ \frac{\Phi _i ^2}{\lambda _i ^2} + \frac{1}{\prod _{k=1} ^{i-1} \sin ^2 \alpha _k}\left( \frac{dS_{\alpha _i}}{d\alpha _i}\right) ^2\right] \, ,\nonumber \\
L_1 ^2&=&\sum_{i=1}^p \left[ \frac{\Psi _i ^2}{\nu _i ^2} + \frac{1}{\prod _{k=1} ^{i-1} \sin ^2 \beta _k}\left( \frac{dS_{\beta _i}}{d\beta _i}\right) ^2\right] \, , \label{albesep}
\end{eqnarray}
where $J_1^2$ and $L_1^2$ are separation constants. Then the remaining terms in the Hamilton-Jacobi equations can be explicitly separated to give ordinary differential equations for $r$ and $\theta$:
\begin{eqnarray}
K=&&m^2r^2- \left[\frac{\Sigma_a\Sigma_b}{\lambda(1-\lambda r^2)}+\frac{2M}{Z(1-\lambda r^2)}+\frac{4M^2}{Z(1-\lambda r^2)^2}\right]E^2 + \frac{V-2M}{Z}\left[\frac{dS_r(r)}{dr}\right]^2\nonumber \\
&+&2\left[\frac{a\Sigma_a\Sigma_b}{(1-\lambda r^2)}+\frac{2M\lambda a}{Z(1-\lambda r^2)^2}+\frac{4M^2a\Sigma_a}{Z(1-\lambda r^2)^2(V-2M)(r^2+a^2)}\right. \nonumber \\
&+&\left.\frac{2Ma}{Z(1-\lambda r^2) (r^2+a^2)}
\right]\sum_{i=1}^m (-E){\Phi_i} \nonumber \\
&+&2\left[\frac{b\Sigma_a\Sigma_b}{(1-\lambda
r^2)}+\frac{2M\lambda
b}{Z(1-\lambda r^2)^2}+\frac{4M^2b\Sigma_b}{Z(1-\lambda r^2)^2(V-2M)(r^2+b^2)}\right.\nonumber \\
&+&\left.\frac{2Mb}{Z(1-\lambda r^2) (r^2+b^2)}
\right]\sum_{j=1}^p (-E){\Psi_j} + \sum_{i=1}^m\sum_{j=1}^m
\left[\lambda^2a^2\left(\frac{\Sigma_a\Sigma_b}{\lambda(1-\lambda
r^2)}+\frac{2M}{Z(1-\lambda r^2)}\right) \right. \nonumber \\
&-& \left. \frac{4M^2a^2\Sigma_a^2}{Z(V-2M)(r^2+a^2)^2}-\frac{Q^{ij}}{Z}\right]\Phi_i \Phi_j \nonumber \\
&+& \sum_{i=1}^p\sum_{j=1}^p
\left[\lambda^2b^2\left(\frac{\Sigma_a\Sigma_b}{\lambda(1-\lambda
r^2)}+\frac{2M}{Z(1-\lambda r^2)}\right) -
\frac{4M^2b^2\Sigma_a^2}{Z(V-2M)(r^2+b^2)^2}\right. \nonumber \\
&-&\left.\frac{Q^{(i+m)(j+m)}}{Z}\right]\Psi_i \Psi_j
+2\sum_{i=1}^m\sum_{j=1}^p
\left[\lambda^2ab\left(\frac{\Sigma_a\Sigma_b}{\lambda(1-\lambda
r^2)}+\frac{2M}{Z(1-\lambda r^2)}\right) \right. \nonumber \\
&-&\left.\frac{4M^2ab\Sigma_a\Sigma_b}{Z(V-2M)(r^2+a^2)(r^2+b^2)}-\frac{Q^{i(j+m)}}{Z}\right]\Phi_i\Psi_j
\nonumber \\
&+& \frac{\Sigma _a (r^2+b^2)}{r^2+a^2}J_1^2 + \frac{\Sigma _b
(r^2+a^2)}{r^2+b^2}L_1^2\,, \label{rsep}
\end{eqnarray}
and
\begin{eqnarray}
-K=&&m^2a^2\cos ^2 \theta + m^2b^2\sin ^2 \theta +
\Delta_\theta\left(\frac{dS_{\theta}}{d\theta}\right)^2 + \Sigma_a \cot ^2
\theta J_1 ^2 + \Sigma_b \tan ^2 \theta L_1 ^2
\nonumber \\
&+&\frac{\Sigma_a\Sigma_b}{\lambda\Delta_\theta}E^2
- 2\sum_{i=1}^m \frac{a\Sigma_a\Sigma_b}{\Delta_\theta} E\Phi_i
- 2\sum_{i=1}^p \frac{b\Sigma_a\Sigma_b}{\Delta_\theta}
E\Psi_i +\sum_{i=1}^{m}\sum_{j=1}^{m}\frac{\lambda^2 a^2 \Sigma_a \Sigma_b}{\Delta_{\theta}}
\Phi_i \Phi_j \nonumber \\
&+& \sum_{i=1}^{p}\sum_{j=1}^{p}\frac{\lambda^2 b^2 \Sigma_a \Sigma_b}{\Delta_{\theta}}
\Psi_i \Psi_j +2 \sum_{i=1}^{m}\sum_{j=1}^{p}\frac{\lambda^2 a b\Sigma_a \Sigma_b}
{\Delta_{\theta}} \Phi_i \Psi_j
\label{thetsep}\, ,
\end{eqnarray}
where $K$ is a separation constant.

In order to show complete separation of the Hamilton-Jacobi
equation, we analyze the $\alpha$ and $\beta$ sectors in
(\ref{albesep}) and demonstrate separation of the individual
$\alpha _i$ and $\beta _j$. The pattern here is that of a
Hamiltonian of non-relativistic classical particles on the unit
$(m-1)$-$\alpha$ and the unit $(p-1)$-$\beta$ spheres, with some
potential dependent on the squares of the $\mu_i$. This can easily
be additively separated following the usual procedure, one angle
at a time, and the pattern continues for all integers $m,p \ge 2$.

The separation has the following inductive form for $k=1,...,m-2$, and $l=1,...,p-2$:
\begin{eqnarray}
\left(\frac{dS_{\alpha _k}}{d\alpha _k}\right)^2 &=& J_k^2 - \frac{J^2_{k+1}}{\sin ^2 \alpha _k} - \frac{\Phi ^2 _{m-k+1}}{\cos ^2 \alpha _k} \, , \nonumber \\
\left(\frac{dS_{\alpha _{m-1}}}{d\alpha _{m-1}}\right)^2 &=& J_{m-1}^2 - \frac{\Phi_1^2}{\sin ^2 \alpha _{m-1}} - \frac{\Phi ^2 _2}{\cos ^2 \alpha _{m-1}} \, , \nonumber \\
\left(\frac{dS_{\beta _l}}{d\beta _l}\right)^2 &=& L_l^2 - \frac{L^2_{l+1}}{\sin ^2 \beta _l} - \frac{\Psi ^2 _{p-l+1}}{\cos ^2 \beta _l} \, , \nonumber \\
\left(\frac{dS_{\beta _{p-1}}}{d\beta _{p-1}}\right)^2 &=& L_{p-1}^2 - \frac{\Psi_1^2}{\sin ^2 \beta _{p-1}} - \frac{\Psi ^2 _2}{\cos ^2 \beta _{p-1}} \, . \label{albesep2}
\end{eqnarray}

Thus, the Hamilton-Jacobi equation in the Kerr-(Anti) de Sitter
rotating black hole background in all dimensions with two sets of
possibly unequal rotation parameters has the general separation
\begin{equation}
S=\frac{1}{2}m^2 l -E\tau + \sum_{i=1} ^m  \Phi_i \varphi _i +
\sum_{i=1}^{p}\Psi_i \varphi _{m+i}+S_r (r) +S_{\theta}(\theta)+\sum
_{i=1} ^{m-1} S_{\alpha _i}(\alpha _i) +\sum _{i=1} ^{p-1}S_{\beta
_i}(\beta _i)\,,
\end{equation}
where the $\alpha _i$  and $\beta _j$ are the spherical polar coordinates on the unit $(m-1)$ and unit $(p-1)$
spheres respectively. $S_r(r)$ can be obtained by quadratures from (\ref{rsep}), $S_{\theta}(\theta)$ by quadratures from (\ref{thetsep}), and the
$S_{\alpha _i}(\alpha _ i)$ and the $S_{\beta_j}(\beta _j)$ again by quadratures from (\ref{albesep2}).

\section{The Equations of Motion}
\subsection{Derivation of the Equations of Motion}

To derive the equations of motion, we will write the separated action $S$ from the Hamilton-Jacobi equation
in the following form:
\bea
S&=&\frac{1}{2}m^2 l -E\tau + \sum_{i=1} ^m \Phi_i \varphi _i + \sum _{i=1}^p \Psi_i \varphi _{m+i} + \int ^r \sqrt{R(r')} dr' + \int ^ {\theta} \sqrt{\Theta(\theta')}d\theta' \nonumber \\
&&\qquad+\sum _{i=1} ^{m-1} \int ^{\alpha _i} \sqrt{A_i (\alpha' _i)}d\alpha ' _i + \sum _{i=1} ^{p-1} \int ^{\beta _i} \sqrt{B_i (\beta' _i)}d\beta ' _i \, ,
\eea
where
\begin{eqnarray}
A_k&=&J_k^2 - \frac{J^2_{k+1}}{\sin ^2 \alpha _k} - \frac{\Phi ^2 _{m-k+1}}{\cos ^2 \alpha _k}\,,\qquad k=1,...,m-2\,, \nonumber \\
A_{m-1}&=&J_{m-1}^2 - \frac{\Phi_1^2}{\sin ^2 \alpha _{m-1}} - \frac{\Phi ^2 _2}{\cos ^2 \alpha _{m-1}} \, , \nonumber \\
B_k&=&L_k^2 - \frac{L^2_{k+1}}{\sin ^2 \beta _k} - \frac{\Psi ^2 _{p-k+1}}{\cos ^2 \beta _k}\,,\qquad k=1,...,p-2\,, \nonumber \\
B_{p-1}&=&L_{p-1}^2 - \frac{\Psi_1^2}{\sin ^2 \beta _{p-1}} - \frac{\Psi ^2 _2}{\cos ^2 \beta _{p-1}} \, ,
\end{eqnarray}
$\Theta$ is obtained from (\ref{thetsep}) as
\begin{eqnarray}
\Delta_\theta \Theta=&-&m^2a^2\cos ^2 \theta - m^2b^2\sin ^2
\theta - \Sigma_a \cot ^2 \theta J_1 ^2 -\Sigma_b \tan ^2 \theta L_1 ^2
-\frac{\Sigma_a\Sigma_b}{\Delta_\theta}E^2 \nonumber \\
&+& 2\sum_{i=1}^m \frac{a\Sigma_a\Sigma_b}{\Delta_\theta} E\Phi_i
+ 2\sum_{i=1}^m \frac{b\Sigma_a\Sigma_b}{\Delta_\theta} E\Psi_i -\sum_{i=1}^{m}\sum_{j=1}^{m}\frac{\lambda^2 a^2 \Sigma_a \Sigma_b}{\Delta_{\theta}}
\Phi_i \Phi_j \nonumber \\
&-& \sum_{i=1}^{p}\sum_{j=1}^{p}\frac{\lambda^2 b^2 \Sigma_a \Sigma_b}{\Delta_{\theta}}
\Psi_i \Psi_j -2 \sum_{i=1}^{m}\sum_{j=1}^{p}\frac{\lambda^2 a b\Sigma_a \Sigma_b}
{\Delta_{\theta}} \Phi_i \Psi_j-K
\end{eqnarray}
and $R$ is obtained from (\ref{rsep}) as
\begin{eqnarray}
\frac{V-2M}{Z}R=&-&m^2r^2+ \left[\frac{\Sigma_a\Sigma_b}{\lambda(1-\lambda r^2)}+\frac{2M}{Z(1-\lambda r^2)}+\frac{4M^2}{Z(1-\lambda r^2)^2}\right]E^2 \nonumber \\
&-&2\left[\frac{a\Sigma_a\Sigma_b}{(1-\lambda r^2)}+\frac{2M\lambda a}{Z(1-\lambda r^2)^2}+\frac{4M^2a\Sigma_a}{Z(1-\lambda r^2)^2(V-2M)(r^2+a^2)}\right. \nonumber \\
&+&\left.\frac{2Ma}{Z(1-\lambda r^2) (r^2+a^2)}
\right]\sum_{i=1}^m (-E){\Phi_i} \nonumber \\
&-&2\left[\frac{b\Sigma_a\Sigma_b}{(1-\lambda
r^2)}+\frac{2M\lambda
b}{Z(1-\lambda r^2)^2}+\frac{4M^2b\Sigma_b}{Z(1-\lambda r^2)^2(V-2M)(r^2+b^2)}\right.\nonumber \\
&+&\left.\frac{2Mb}{Z(1-\lambda r^2) (r^2+b^2)}
\right]\sum_{j=1}^p (-E){\Psi_j} - \sum_{i=1}^m\sum_{j=1}^m
\left[\lambda^2a^2\left(\frac{\Sigma_a\Sigma_b}{\lambda(1-\lambda
r^2)} \right. \right.\nonumber \\
&+& \left. \left.\frac{2M}{Z(1-\lambda r^2)}\right) -\frac{4M^2a^2\Sigma_a^2}{Z(V-2M)(r^2+a^2)^2}-\frac{Q^{ij}}{Z}\right]\Phi_i \Phi_j \nonumber \\
&-& \sum_{i=1}^p\sum_{j=1}^p
\left[\lambda^2b^2\left(\frac{\Sigma_a\Sigma_b}{\lambda(1-\lambda
r^2)}+\frac{2M}{Z(1-\lambda r^2)}\right) -
\frac{4M^2b^2\Sigma_a^2}{Z(V-2M)(r^2+b^2)^2}\right. \nonumber \\
&-&\left.\frac{Q^{(i+m)(j+m)}}{Z}\right]\Psi_i \Psi_j
-2\sum_{i=1}^m\sum_{j=1}^p
\left[\lambda^2ab\left(\frac{\Sigma_a\Sigma_b}{\lambda(1-\lambda
r^2)}+\frac{2M}{Z(1-\lambda r^2)}\right) \right. \nonumber \\
&-&\left.\frac{4M^2ab\Sigma_a\Sigma_b}{Z(V-2M)(r^2+a^2)(r^2+b^2)}-\frac{Q^{i(j+m)}}{Z}\right]\Phi_i\Psi_j\nonumber \\
&-& \frac{\Sigma _a (r^2+b^2)}{r^2+a^2}J_1^2 - \frac{\Sigma _b
(r^2+a^2)}{r^2+b^2}L_1^2+K \, .
\end{eqnarray}

To obtain the equations of motion, we differentiate $S$ with respect to the
parameters $m^2,K,E,J_i^2,L^2_j,\Phi_i,\Psi_j$ and set these derivatives to equal
other constants of motion. However, we can set all these new constants of motion
to zero (following from freedom in choice of origin for the corresponding
coordinates, or alternatively by changing the constants of integration). Following this procedure, we get the following equations of motion:
\vspace{.5cm}
\begin{eqnarray}
\frac{\pa S}{\pa m^2}&=&0 \Rightarrow l = \int \frac{Z r^2}{V-2M} \frac{dr}{\sqrt{R}} + \int \frac{(a^2\cos ^2 \theta + b^2 \sin ^2 \theta) d\theta}{\Delta_\theta\sqrt{\Theta}} \, , \nonumber \\
\frac{\pa S}{\pa K}&=&0 \Rightarrow \int \frac{d\theta}{\Delta_\theta\sqrt{\Theta}} = \int \frac{Z}{V-2M} \frac{dr}{\sqrt{R}} \, , \nonumber \\
\frac{\pa S}{\pa J_1^2}&=&0 \Rightarrow \int \frac{d\alpha _1}{\sqrt{A_1}} = \int \frac{Z}{V-2M} \frac{\Sigma_a(r^2+b^2)}{r^2+a^2} \frac{dr}{\sqrt{R}} + \int  \frac{\Sigma_a \cot ^2 \theta d\theta}{\Delta_\theta\sqrt{\Theta}} \, , \nonumber \\
\frac{\pa S}{\pa J_k^2}&=&0 \Rightarrow \int \frac{d\alpha _k}{\sqrt{A_k}} = \int \frac{1}{\sin ^2 \alpha _{k-1}} \frac{d\alpha _{k-1}}{\sqrt{A_{k-1}}} \, , \qquad k=2,...,m-2 \, , \nonumber \\
\frac{\pa S}{\pa L_1^2}&=&0 \Rightarrow \int \frac{d\beta _1}{\sqrt{B_1}} = \int \frac{Z}{V-2M} \frac{\Sigma_b(r^2+a^2)}{r^2+b^2} \frac{dr}{\sqrt{R}} + \int \frac{\Sigma_b \tan ^2 \theta d\theta}{\Delta_\theta\sqrt{\Theta}} \, , \nonumber \\
\frac{\pa S}{\pa L_l^2}&=&0 \Rightarrow \int \frac{d\beta _l}{\sqrt{B_l}} = \int \frac{1}{\sin ^2 \beta _{k-1}} \frac{d\beta _{l-1}}{\sqrt{B_{l-1}}} \, , \qquad l=2,...,p-2 \,. \label{inteqs}
\end{eqnarray}
We can obtain $N$ more equations of motion for the variables $\varphi_i$ by differentiating $S$ with respect to the angular momenta $\Phi_i$ and $\Psi_j$. Another equation can also be obtained by differentiating $S$ with respect to $E$ involving the time coordinate $\tau$. However, these equations are not particularly illuminating, but can be written out
explicitly if necessary by following this procedure. It is often more convenient to rewrite these in the form of first-order differential equations obtained from (\ref{inteqs}) by direct differentiation with respect to the affine parameter. We only list the most relevant ones here:
\begin{eqnarray}
\rho ^2 \frac{dr}{dl} &=& \frac{V-2M}{Z} \sqrt{R} \, , \nonumber \\
\rho ^2 \frac{d\theta}{dl} &=& \Delta_\theta \sqrt{\Theta} \, , \nonumber \\
\frac{(r^2+a^2)}{\Sigma_a} \frac{d\alpha _k}{dl} &=& \frac{\sqrt{A_k}}{\sin ^2 \theta \prod _{i=1}^{k-1} \sin ^2 \alpha _i} \, , \qquad k=1,...,m-1 \, ,\nonumber \\
\frac{(r^2+b^2)}{\Sigma_b} \frac{d\beta _k}{dl } &=&
\frac{\sqrt{B_l}}{\cos ^2 \theta \prod _{i=1}^{l-1} \sin ^2
\beta_i} \, , \qquad l=1,...,p-1 \, , \label{eqns}
\end{eqnarray}

\subsection{Analysis of the Radial Equation}
Worldlines of particles in these backgrounds are completely
specified by the values of the conserved quantities $E,K,L^2_i,
J^2 _j$, and by the initial values of the coordinates. We will
consider particle motion in the black hole exterior. Allowed
regions of particle motion necessarily need to have positive value
for the quantity $R$, owing to equation (\ref{eqns}). We determine
some of the possibilities of the allowed motion.

At large radius $r$, the dominant contribution to $R$, in the case
of $\lambda =0$, is $E^2 -m^2$. Thus we can say that for
$E^2<m^2$, we cannot have unbounded orbits, whereas for $E^2>m^2$,
such orbits are possible. For the case of nonzero $\lambda$, the
dominant term at large $r$ in $R$ (or rather the slowest decaying
term) is $\frac{m^2}{\lambda r^2}$. Thus in the case of the
Kerr-Anti-de Sitter background, only bound orbits are possible,
whereas in the Kerr-de Sitter backgrounds, both unbounded and
bound orbits may be possible.

In order to study the radial motion of particles in these metrics, it is useful
to cast the radial equation of motion into a different form. Decompose $R$ as a
quadratic in $E$ as follows:
\begin{equation}
R=\alpha E^2-2\beta E + \gamma \,,
\end{equation}
where
\begin{eqnarray}
\alpha &=& \frac{Z}{V-2M} \left[ \frac{\Sigma_a\Sigma_b}{\lambda(1-\lambda r^2)}+\frac{2M}{Z(1-\lambda r^2)}+\frac{4M^2}{Z(1-\lambda r^2)^2}\right] \, , \nonumber \\
\beta &=& \frac{-Z}{V-2M}\left[\frac{a\Sigma_a\Sigma_b}{(1-\lambda r^2)}+\frac{2M\lambda a}{Z(1-\lambda r^2)^2}+\frac{4M^2a\Sigma_a}{Z(1-\lambda r^2)^2(V-2M)(r^2+a^2)}\right. \nonumber \\
&+&\left.\frac{2Ma}{Z(1-\lambda r^2) (r^2+a^2)}
\right]\sum_{i=1}^m {\Phi_i} \nonumber \\
&-&\frac{Z}{V-2M}\left[\frac{b\Sigma_a\Sigma_b}{(1-\lambda
r^2)}+\frac{2M\lambda
b}{Z(1-\lambda r^2)^2}+\frac{4M^2b\Sigma_b}{Z(1-\lambda r^2)^2(V-2M)(r^2+b^2)}\right.\nonumber \\
&+&\left.\frac{2Mb}{Z(1-\lambda r^2) (r^2+b^2)}
\right]\sum_{j=1}^p {\Psi_j} \, ,\nonumber \\
\gamma &=& \left\{-\sum_{i=1}^m\sum_{j=1}^m
\left[\lambda^2a^2\left(\frac{\Sigma_a\Sigma_b}{\lambda(1-\lambda
r^2)} + \frac{2M}{Z(1-\lambda r^2)}\right) -\frac{4M^2a^2\Sigma_a^2}{Z(V-2M)(r^2+a^2)^2}-\frac{Q^{ij}}{Z}\right]\Phi_i \Phi_j \right. \nonumber \\
&-& \left.\sum_{i=1}^p\sum_{j=1}^p
\left[\lambda^2b^2\left(\frac{\Sigma_a\Sigma_b}{\lambda(1-\lambda
r^2)}+\frac{2M}{Z(1-\lambda r^2)}\right) -
\frac{4M^2b^2\Sigma_a^2}{Z(V-2M)(r^2+b^2)^2}\right. \right. \nonumber \\
&-&\left.\left.\frac{Q^{(i+m)(j+m)}}{Z}\right]\Psi_i \Psi_j
-2\sum_{i=1}^m\sum_{j=1}^p
\left[\lambda^2ab\left(\frac{\Sigma_a\Sigma_b}{\lambda(1-\lambda
r^2)}+\frac{2M}{Z(1-\lambda r^2)}\right) \right. \right. \nonumber \\
&-&\left.\left.\frac{4M^2ab\Sigma_a\Sigma_b}{Z(V-2M)(r^2+a^2)(r^2+b^2)}-\frac{Q^{i(j+m)}}{Z}\right]\Phi_i\Psi_j\right.\nonumber \\
&-& \left.\frac{\Sigma _a (r^2+b^2)}{r^2+a^2}J_1^2 - \frac{\Sigma
_b (r^2+a^2)}{r^2+a^2}L_1^2+K -m^2r^2\right\}\frac{Z}{V-2M}\, .
\end{eqnarray}

The turning points for trajectories in the radial motion (defined by the condition
$R=0$) are given by $E=V_{\pm}$ where
\be
V_{\pm} = \frac{\beta \pm \sqrt{\beta ^2 -\alpha \gamma }}{\alpha} \,.
\ee
These functions, called the effective potentials \cite{frolov1},
determine allowed regions of motion. In this form, the radial equation
is much more suitable for detailed numerical analysis for specific values of
parameters.

\subsection{Analysis of the Angular Equations}
Another class of interesting motions possible describes motion at
a constant value of $\alpha _i$ or $\beta_j$. These are analogous
to the same class of motions analyzed in \cite{VSP2}. We briefly
summarize them here. These motions are described by the
simultaneous equations
\begin{equation}
A_i (\alpha_i= \alpha _{i0})=\frac{dA _i}{d\alpha _i}(\alpha _i=\alpha _{i0}) =0 \,, \qquad i=1,...,m-1\,,
\end{equation}
in the case of constant $\alpha_i$ motion, where $\alpha _{i0}$ is the constant value of $\alpha _i$ along this trajectory, or by the simultaneous equations
\begin{equation}
B_i (\beta_i= \beta _{i0})=\frac{dB _i}{d\beta _i}(\beta _i=\beta _{i0}) =0 \,, \qquad i=1,...,p-1\,,
\end{equation}
in the case of constant $\beta_i$ motion, where $\beta _{i0}$ is the constant value of $\beta _i$ along this trajectory.

These equations can be explicitly solved. In the case of constant $\alpha _i$ motion, we get the relations
\begin{eqnarray}
\frac{J^2_{i+1}}{\sin ^4 \alpha _i} &=& \frac{\Phi^2 _{m-i-1}}{\cos ^4 \alpha _i} \,,\nonumber \\
J^2_i&=&\frac{J^2_{i+1}}{\sin ^2 \alpha _i} +\frac{\Phi^2_{m-i+1}}{\cos ^2 \alpha_i}\,,\qquad i=1,...,m-1\,.
\end{eqnarray}
Note that if $\alpha_{i0}=0$, then $J^2 _{i+1} =0$, and if $\alpha _{i0} = \pi /2$, then $\Phi_{m-i+1} ^2=0$. Similarly, in the case of constant $\beta_i$ motion, we get the relations
\begin{eqnarray}
\frac{L^2_{i+1}}{\sin ^4 \beta _i} &=& \frac{\Psi^2 _{p-i-1}}{\cos ^4 \beta _i} \,,\nonumber \\
L^2_i&=&\frac{L^2_{i+1}}{\sin ^2 \beta _i} +\frac{\Psi^2_{p-i+1}}{\cos ^2 \beta_i}\,,\qquad i=1,...,p-1\,.
\end{eqnarray}
Again if $\beta_{i0}=0$, then $L^2 _{i+1} =0$, and if $\beta _{i0} = \pi /2$, then $\Psi_{p-i+1} ^2=0$.

Examining $A_k $ in the general case, $\alpha _k=0$ can only be reached if $J_{k+1}=0$, and
$\alpha _k =\pi/2$ can be only be reached if $\Phi_{m-k+1}=0$. The orbit will completely be in the subspace
$\alpha _ k=0$ only if $J_k ^2 = \Phi^2 _{m-k+1}$ and will completely be in the subspace $\alpha _k =\pi/2$ only if $J_k ^2 = J_{k+1} ^2$. Analogous results hold for constant $\beta _i$ motion.

Again these equations are in a form suitable for numerical analysis for specific values of the black hole and particle parameters.

\section{Dynamical Symmetry}
The spacetimes discussed here are stationary and ``axisymmetric";
i.e., $\partial / \partial \tau$ and $\partial / \partial \varphi
_i$ are Killing vectors and have associated conserved quantities,
$-E$, $\Phi_i$, and $\Psi_i$. In general if $\eta$ is a Killing
vector, then $\eta ^{\mu} p_{\mu}$ is a conserved quantity, where
$p$ is the momentum. Note that this quantity is first order in the
momenta.

In the case of only two sets of possibly unequal rotation
parameters, more Killing vectors exist since the spacetime
acquires additional dynamical symmetry. We have complete symmetry
between the various planes of rotation characterized by the same
value of rotation parameter $a_i=a$, and we can ``rotate" one into
another. Similarly, we have symmetry between the planes of
rotation characterized by the same value of the rotation parameter
$a_i=b$, and we can ``rotate" these into one another as well. The
vectors that generate these transformations are the required
Killing vectors. The explicit construction of such Killing vectors
is done in \cite{VSP1}. In this case, we get two independent sets
of such Killing vectors, associated with the constant $a$ and $b$
value rotations.

In  an odd number of spacetime dimensions, if $a\neq b$ and both
are nonvanishing, then the rotational symmetry group is
$U(m)\times U(p)$. If one of them is zero, but the other is
nonzero (we take the nonzero one to be $a$), then the rotational
symmetry group is $U(m)\times O(2p)$. In the case when $a=b\neq0$,
the rotational symmetry group is $U(m+p)$. In the case when
$a=b=0$, i.e. in the Schwarzschild metric, the rotational symmetry
group is $O(2m+2p)$. In an even number of spacetime dimensions,
$b=0$ in the cases we have analyzed. If $a\neq0$, then the
rotational symmetry group is $U(m)\times O(2p-1)$, and in the case
when $a=b=0$, i.e. in the Schwarzschild metric, the rotational
symmetry group is $O(2m+2p-1)$. Note that since these metrics are
stationary, the full dynamical symmetry group is the direct
product of $\mathbf{R}$ and the rotational symmetry group, where
$\mathbf{R}$ is the additive group of real numbers parameterizing
$\tau$.

We also obtain a non-trivial irreducible second-order Killing
tensor, whose existence is the principal reason that permits the
separation of the $r-\theta$ equations. This Killing tensor is a
generalization of the result obtained in the five dimensional case
in \cite{frolov1}. This is obtained from the separation constant
$K$ in (\ref{rsep}) and (\ref{thetsep}). We choose to analyze the
latter.
\begin{eqnarray}
K=&-&m^2a^2\cos ^2 \theta -m^2 b^2 \sin ^2\theta
-\frac{\Sigma_a\Sigma_b}{\lambda\Delta_\theta}E^2- \Sigma_a \cot ^2 \theta
J_1 ^2 - \Sigma_b \tan ^2 \theta L_1 ^2 - \Delta_\theta\left(\frac{\partial
S}{\partial \theta}\right)^2 \nonumber \\
&+&2\sum_{i=1}^m\frac{a\Sigma_a\Sigma_b}{\Delta_\theta}E\Phi_i+2\sum_{j=1}^p\frac{b\Sigma_a\Sigma_b}{\Delta_\theta}E\Psi_i-
\sum_{i=1}^{m}\sum_{j=1}^{m}\frac{\lambda^2 a^2 \Sigma_a \Sigma_b}{\Delta_{\theta}}
\Phi_i \Phi_j \nonumber \\
&-& \sum_{i=1}^{p}\sum_{j=1}^{p}\frac{\lambda^2 b^2 \Sigma_a \Sigma_b}{\Delta_{\theta}}
\Psi_i \Psi_j -2 \sum_{i=1}^{m}\sum_{j=1}^{p}\frac{\lambda^2 a b\Sigma_a \Sigma_b}
{\Delta_{\theta}} \Phi_i \Psi_j\,.
\end{eqnarray}
The Killing tensor $K^{\mu \nu}$ is obtained from this separation constant (which is quadratic in the canonical momenta) using the relation $K=K^{\mu \nu} p_{\mu} p_{\nu}$. Its is then easy to see that
\begin{eqnarray}
K^{\mu \nu} = &-&g^{\mu\nu}\left(a^2\cos ^2 \theta + b^2 \sin ^2
\theta \right)
-\frac{\Sigma_a\Sigma_b}{\lambda\Delta_\theta}\delta
^{\mu}_{\tau}\delta ^{\nu}_{\tau}- \Sigma_a \cot ^2 \theta J_1 ^{\mu \nu} -
\Sigma_b \tan ^2 \theta L_1 ^{\mu \nu} -\Delta_\theta \delta
^{\mu}_{\theta}\delta ^{\nu}_{\theta} \nonumber \\
&-&\sum_{i=1}^m\frac{a\Sigma_a\Sigma_b}{\Delta_\theta}(\delta
^{\mu}_{\tau}\delta ^{\nu}_{\varphi_i}+\delta ^{\mu}_{\varphi_i}\delta
^{\nu}_{\tau})-\sum_{j=1}^p\frac{b\Sigma_a\Sigma_b}{\Delta_\theta}(\delta
^{\mu}_{\tau}\delta ^{\nu}_{\varphi_{i+m}}+\delta
^{\mu}_{\varphi_{i+m}}\delta ^{\nu}_{\tau}) \nonumber \\
&-&\sum_{i=1}^{m}\sum_{j=1}^{m}\frac{\lambda^2 a^2 \Sigma_a \Sigma_b}{\Delta_{\theta}}
 \delta^{\mu}_{\varphi_i}\delta ^{\nu}_{\varphi_j} - \sum_{i=1}^{p}\sum_{j=1}^{p}\frac{\lambda^2 b^2 \Sigma_a \Sigma_b}{\Delta_{\theta}}
 \delta^{\mu}_{\varphi_{i+m}}\delta ^{\nu}_{\varphi_{j+m}} \nonumber \\
&-& \sum_{i=1}^{m}\sum_{j=1}^{p}\frac{\lambda^2 a b\Sigma_a \Sigma_b}{\Delta_{\theta}}  (\delta^{\mu}_{\varphi_{i}}\delta ^{\nu}_{\varphi_{j+m}}+\delta^{\mu}_{\varphi_{j+m}}\delta ^{\nu}_{\varphi_{i}})\,.
\end{eqnarray}
where $J_1^{\mu \nu}$ and $L_1^{\mu
\nu}$ are the reducible Killing tensors associated with the
$\alpha$ and $\beta$ separation.

The existence of these additional Killing vectors and the
nontrivial irreducible Killing tensor, is the principal reason
behind the complete separation of the Hamilton-Jacobi equation.
The nontrivial Killing tensor, in particular, exists due to the
detailed structure of the metrics under consideration and is a
surprising result.

\section{The Scalar Field Equation}
Consider a scalar field $\Psi$ in a gravitational background with the action
\begin{equation}
S[\Psi]=-\frac{1}{2}\int d^Dx\sqrt{-g}((\nabla \Psi)^2+ \alpha R \Psi ^2 + m^2
\Psi ^2 ) \,,
\end{equation}
where we have included a curvature dependent coupling. However, in
the Kerr-(Anti) de Sitter background, $R = \lambda$ is constant.
As a result we can trade off the curvature coupling for a
different mass term. So it is sufficient to study the massive
Klein-Gordon equation in this background. We will simply set
$\alpha=0$ in the following. Variation of the action leads to the
Klein-Gordon equation
\begin{equation}
\frac{1}{\sqrt{-g}}\partial _{\mu}(\sqrt{-g} g^{\mu \nu}\partial _{\nu} \Psi
)=m^2 \Psi \,.\label{KG1}
\end{equation}

As discussed by Carter \cite{Carter}, the assumption of
separability of the Klein-Gordon equation usually
implies separability of the Hamilton-Jacobi equation. Conversely, if the
Hamilton-Jacobi equation does not separate, the Klein-Gordon equation seems
unlikely to separate. We can also see this explicitly (as in the case of the Hamilton-Jacobi equation), since the $(r,\tau,\phi_i)$ sector
has coefficients in the equations that explicitly depend on the $\mu_i$ except when of all $a_i=a$, in which case separation seems likely. We will again consider the much more general case of
two sets of possibly unequal sets of rotation parameters $a$ and $b$. We continue using the same numbering conventions for the variables.

Once again, we impose the constraint (\ref{constraint}) and decompose the
$\mu_i$ in two sets of spherical polar coordinates as in (\ref{mudef}) and (\ref{lnsphere}). We calculate
the determinant of the metric to be
\begin{eqnarray}
g&=&\frac{-r^2 \rho ^4 (r^2+a^2)^{2m-2}(r^2+b^2)^{2p-2-\epsilon}}{\Sigma_a^{2m} \Sigma_b ^{2p-2\epsilon}}\sin ^{4m-2}\theta \cos ^{4p-2-2\epsilon}\theta  \nonumber \\
&&\, \star \left[ \prod _{j=1}^{m-1} \sin ^{4m-4j-2}\alpha _j \cos
^2 \alpha _j \right]\left[ \prod _{k=1}^{p-1} \sin ^{4p-4k-2}\beta
_k \cos ^2 \beta _k \right] \cos ^{-2\epsilon} \beta _1 \,.
\end{eqnarray}
For convenience we write $g=-\frac{RTAB\rho ^4}{\Sigma_a^{2m}
\Sigma_b ^{2p-2\epsilon}}$, where
\begin{eqnarray}
R&=&r^2(r^2+a^2)^{2m-2}(r^2+b^2)^{2p-2-\epsilon} \, ,\nonumber \\
T&=&\sin^{4m-2}\theta \cos ^{4p-2-2\epsilon} \theta \, ,\nonumber \\
A&=&\prod _{j=1}^{m-1} \sin ^{4m-4j-2}\alpha _j \cos ^2 \alpha _j \, ,\nonumber \\
B&=&\prod _{k=1}^{p-1} \sin ^{4p-4k-2}\beta _k \cos ^2 \beta _k \cos ^{-2\epsilon} \beta _1 \, .
\end{eqnarray}
Note that $R$ and $T$ are functions of $r$ and $\theta$ only, and $A$ and $B$ only depend on the set of variables $\alpha _i$ and $\beta _j$ respectively.
Then the Klein-Gordon equation in this background (\ref{KG1}) becomes
\begin{eqnarray}
m^2 \Psi &=& \frac{1}{\rho ^2 \sqrt {R}} \partial _r
\left(\sqrt{R} \frac{V-2M}{Z} \partial _r \Psi \right) +
\frac{\Sigma_a}{(r^2+a^2)\sin^2\theta}\sum_{i=1} ^{m}
\frac{1}{\lambda _i^2}\partial_{\varphi_i} ^2 \Psi  \nonumber \\
&+& \frac{\Sigma_b}{(r^2+b^2)\cos^2\theta}\sum_{i=1} ^{p}
\frac{1}{\nu _{i} ^2}\partial _{\varphi_{i+m}} ^2 \Psi
+\frac{1}{\rho^2\sqrt {T}} \partial _{\theta} \left( \sqrt{T}
\Delta_{\theta} \partial _ {\theta} \Psi\right) \nonumber \\
&+& \frac{\Sigma_a}{(r^2+a^2)\sin ^2 \theta} \left[\sum _{i=1} ^{m-1} \frac{1}{\sqrt{A}} \partial _{\alpha_i} \left( \frac{\sqrt{A}}{\prod _{k=1} ^ {i-1} \sin ^2 \alpha _k } \partial _{\alpha _i} \Psi\right)\right] \nonumber \\
&+&\frac{\Sigma_b}{(r^2+b^2)\cos ^2 \theta} \left[\sum
_{i=1}^{p-1} \frac{1}{\sqrt{B}} \partial _{\beta_i} \left(
\frac{\sqrt{B}}{\prod _{k=1} ^ {i-1} \sin ^2 \beta _k } \partial
_{\beta _i} \Psi \right)\right] \nonumber \\
&+& \left[\frac{\Sigma_a \Sigma _b}{\lambda\rho ^2
\Delta_\theta}-\frac{\Sigma_a\Sigma_b}{\rho^2\lambda(1-\lambda
r^2)}-\frac{2M}{\rho ^2 Z(1-\lambda
r^2)}-\frac{4M^2}{\rho^2Z(1-\lambda
r^2)^2}\right]\partial_{\tau}^2 \Psi \nonumber \\
&+&2\left[\frac{a\Sigma_a \Sigma _b}{\rho ^2 \Delta_\theta}-\frac{a\Sigma_a\Sigma_b}{\rho^2(1-\lambda r^2)}-\frac{2M\lambda a}{\rho ^2 Z(1-\lambda r^2)^2}-\frac{4M^2a\Sigma_a}{\rho^2Z(1-\lambda r^2)^2(V-2M)(r^2+a^2)}\right. \nonumber \\
&-&\left.\frac{2Ma}{\rho^2Z(1-\lambda r^2) (r^2+a^2)} \right]\sum_{i=1}^m \partial^2_{\tau \varphi_i}\Psi \nonumber \\
&+&2\left[\frac{b\Sigma_a \Sigma _b}{\rho ^2 \Delta_\theta}-\frac{b\Sigma_a\Sigma_b}{\rho^2(1-\lambda r^2)}-\frac{2M\lambda b}{\rho ^2 Z(1-\lambda r^2)^2}-\frac{4M^2b\Sigma_b}{\rho^2Z(1-\lambda r^2)^2(V-2M)(r^2+b^2)}\right. \nonumber \\
&-&\left.\frac{2Mb}{\rho^2Z(1-\lambda r^2) (r^2+b^2)} \right]\sum_{j=1}^{p} \partial^2_{\tau \varphi_{m+j}}\Psi  \nonumber \\
&+& \sum_{i=1}^m\sum_{j=1}^m
\left[\lambda^2a^2\left(\frac{\Sigma_a \Sigma _b}{\rho ^2
\Delta_\theta}-\frac{\Sigma_a\Sigma_b}{\rho^2\lambda(1-\lambda
r^2)}-\frac{2M}{\rho ^2 Z(1-\lambda r^2)}\right) \right. \nonumber \\
&+& \left. \frac{4M^2a^2\Sigma_a^2}{\rho^2
Z(V-2M)(r^2+a^2)^2}+\frac{Q^{ij}}{\rho^2Z}\right]\partial ^2 _{\varphi_i \varphi_j}\Psi \nonumber \\
&+& \sum_{i=1}^p\sum_{j=1}^p
\left[\lambda^2b^2\left(\frac{\Sigma_a \Sigma _b}{\rho ^2
\Delta_\theta}-\frac{\Sigma_a\Sigma_b}{\rho^2\lambda(1-\lambda
r^2)}-\frac{2M}{\rho ^2 Z(1-\lambda r^2)}\right) \right. \nonumber \\
&+& \left. \frac{4M^2b^2\Sigma_a^2}{\rho^2
Z(V-2M)(r^2+b^2)^2}+\frac{Q^{(i+m)(j+m)}}{\rho^2Z}\right]\partial ^2 _{\varphi_{i+m} \varphi_{j+m}}\Psi \nonumber \\
&+& 2\sum_{i=1}^m\sum_{j=1}^p
\left[\lambda^2ab\left(\frac{\Sigma_a \Sigma _b}{\rho ^2
\Delta_\theta}-\frac{\Sigma_a\Sigma_b}{\rho^2\lambda(1-\lambda
r^2)}-\frac{2M}{\rho ^2 Z(1-\lambda r^2)}\right) \right. \nonumber \\
&+& \left. \frac{4M^2ab\Sigma_a\Sigma_b}{\rho^2
Z(V-2M)(r^2+a^2)(r^2+b^2)}+\frac{Q^{i(j+m)}}{\rho^2Z}\right]\partial
^2 _{\varphi_{i} \varphi_{j+m}}\Psi \, .
\end{eqnarray}
We attempt the usual multiplicative separation for $\Psi$ in the following form:
\begin{equation}
\Psi=\Phi_r (r) \Phi_{\theta}(\theta)e^{-iE\tau}e^{i\sum_i^m
\Phi_i \varphi _i} e^{i\sum_i^p \Psi_i \varphi
_{m+i}}\left(\prod_{i=1}^{m-1}\Phi_{\alpha_i}(\alpha_i)\right)\left(\prod_{i=1}^{p-1}\Phi_{\beta_i}(\beta_i)\right)\,,\label{KGanz}
\end{equation}
where we again adopt the convention that $\Psi_p=0$ in the case of
even dimensional spacetimes.

 The Klein-Gordon equation then completely separates. The $r$ and $\theta$ equations are given as
\begin{eqnarray}
K&=&\frac{1}{\Phi_r \sqrt{R}} \frac{d}{dr}\left(\sqrt{R}
\frac{V-2M}{Z} \frac{d\Phi _r}{dr}\right)+ \left[\frac{\Sigma_a\Sigma_b}{\lambda(1-\lambda r^2)}+\frac{2M}{Z(1-\lambda r^2)}+\frac{4M^2}{Z(1-\lambda r^2)^2}\right]E^2 \nonumber \\
&+& \sum_{i=1}^m\sum_{j=1}^m
\left[\lambda^2a^2\left(\frac{\Sigma_a\Sigma_b}{\lambda(1-\lambda
r^2)}+\frac{2M}{Z(1-\lambda r^2)}\right)
-  \frac{4M^2a^2\Sigma_a^2}{Z(V-2M)(r^2+a^2)^2}-\frac{Q^{ij}}{Z}\right]\Phi_i\Phi_j \nonumber \\
&+& \sum_{i=1}^p\sum_{j=1}^p
\left[\lambda^2b^2\left(\frac{\Sigma_a\Sigma_b}{\lambda(1-\lambda r^2)}+\frac{2M}{Z(1-\lambda r^2)}\right) \right. \nonumber \\
&-& \left. \frac{4M^2b^2\Sigma_a^2}{Z(V-2M)(r^2+b^2)^2}-\frac{Q^{(i+m)(j+m)}}{Z}\right]\Psi_i\Psi_j \nonumber \\
&+& 2\sum_{i=1}^m\sum_{j=1}^p
\left[\lambda^2ab\left(\frac{\Sigma_a\Sigma_b}{\lambda(1-\lambda
r^2)}+\frac{2M}{ Z(1-\lambda r^2)}\right) \right. \nonumber \\
&-&\left.\frac{4M^2ab\Sigma_a\Sigma_b}{Z(V-2M)(r^2+a^2)(r^2+b^2)}-\frac{Q^{i(j+m)}}{Z}\right]
\Phi_i \Psi_j \nonumber \\
&-&2\left[\frac{a\Sigma_a\Sigma_b}{(1-\lambda r^2)}+\frac{2M\lambda a}{Z(1-\lambda r^2)^2}+\frac{4M^2a\Sigma_a}{Z(1-\lambda r^2)^2(V-2M)(r^2+a^2)}\right. \nonumber \\
&+&\left.\frac{2Ma}{Z(1-\lambda r^2) (r^2+a^2)} \right]\sum_{i=1}^m E\Phi_i \nonumber \\
&-&2\left[\frac{b\Sigma_a\Sigma_b}{(1-\lambda r^2)}+\frac{2M\lambda b}{Z(1-\lambda r^2)^2}+\frac{4M^2b\Sigma_b}{Z(1-\lambda r^2)^2(V-2M)(r^2+b^2)}\right. \nonumber \\
&+&\left.\frac{2Mb}{Z(1-\lambda r^2) (r^2+b^2)}
\right]\sum_{j=1}^{p} E\Psi_j -\Sigma_a\frac{r^2+b^2}{r^2+a^2}\sum_{i=1}^m K_1 \Phi^2_i \nonumber \\
&-& \Sigma_b\frac{r^2+a^2}{r^2+b^2}\sum_{j=1}^p M_1 \Psi^2_j -m^2 r^2 \, ,
\end{eqnarray}
and
\begin{eqnarray}
-K&=& \frac{1}{\Phi_{\theta}\sqrt{T}}\frac{d}{d\theta}\left(\sqrt{T} \Delta_\theta \frac{d\Phi_{\theta}}{d\theta}\right) -\frac{\Sigma_a\Sigma_b}{\lambda\Delta_\theta}E^2-m^2(a^2\cos ^2 \theta + b^2 \sin ^2 \theta) \nonumber \\
&+& K_1 \cot ^2 \theta + M_1 \tan ^2 \theta -2\lambda
a^2\frac{\Sigma_a\Sigma_b}{\Delta_\theta}\sum_{i=1}^m\sum_{j=1}^m\Phi_i\Phi_j-2\lambda
b^2\frac{\Sigma_a\Sigma_b}{\Delta_\theta}\sum_{i=1}^p\sum_{j=1}^p\Psi_i\Psi_j
\nonumber \\
&-&4\lambda ab \frac{\Sigma_a \Sigma_b}{\Delta_\theta}
\sum_{i=1}^m \sum_{j=1}^p \Phi_i \Psi_j +2 \frac{a \Sigma_a \Sigma_b}{\Delta_{\theta}}\sum_{i=1}^m E \Phi_i +2 \frac{b \Sigma_a \Sigma_b}{\Delta_{\theta}}\sum_{j=1}^p E \Phi_j\, ,
\end{eqnarray}
where $K$, $K_1$ and $M_1$ are separation constants. $K_1$ and $M_1$ encode all the $\alpha$ and $\beta$ dependence respectively and are defined explicitly as follows:
\begin{eqnarray}
K_1 &=& \sum _{i=1} ^{k-1} A_i + \frac{ K_k }{\prod _ {j=1} ^{k-1} \sin ^2\alpha _j }\,, \quad k=1,...,m-1\,,
\end{eqnarray}
where
\begin{eqnarray}
A_i &=&  \frac{1}{\Phi _{\alpha _i} \cos \alpha _i \sin
^{2m-2i-1} \alpha _i \prod _{k=1} ^{i-1} \sin ^2 \alpha _ k}
\frac{d}{d\alpha _i} \left( \cos \alpha _i \sin ^{2m-2i-1} \alpha _i \frac{d\Phi
_{\alpha _i}}{d\alpha _i} \right) \nonumber \\ && -\frac{\Phi^2 _{m-i+1}}{\cos ^2 \alpha _i \prod _{j=1} ^{i-1} \sin
^2{\alpha _j}}  \,,
\end{eqnarray}
and
\begin{eqnarray}
M_1 &=& \sum _{i=1} ^{k-1} B_i + \frac{ M_k }{\prod _ {j=1} ^{k-1} \sin ^2\beta _j }\,, \quad k=1,...,p-1\,,
\end{eqnarray}
and where
\begin{eqnarray}
B_i &=&  \frac{1}{\Psi _{\beta _i} \cos \beta _i \sin
^{2p-2i-1} \beta _i \prod _{k=1} ^{i-1} \sin ^2 \beta _ k}
\frac{d}{d\beta _i} \left( \cos \beta _i \sin ^{2p-2i-1} \beta _i \frac{d\Phi
_{\beta _i}}{d\beta _i} \right) \nonumber \\ && -\frac{\Psi^2 _{p-i+1}}{\cos ^2 \beta _i \prod _{j=1} ^{i-1} \sin
^2{\beta _j}}  \,,
\end{eqnarray}
Then we inductively have the complete separation of the $\alpha _i$ dependence as
\begin{equation}
K_k  = \frac{K _{k+1}}{\sin ^2 \alpha _k} -\frac{\Phi^2 _{n-k+1}}{\cos ^2
\alpha _k} + \frac{1}{\Phi _{\alpha_k} \cos \alpha _k \sin ^{2m-2k-1} \alpha_k}
\frac{d}{d\alpha _k} \left( \cos\alpha _k \sin \alpha _k \frac{d\Phi _{\alpha
_k}}{d\alpha _k} \right) \,,
\end{equation}
where $k=1,...,m-1$, and we use the convention $K_m = -\Phi_1 ^2$. Similarly, the complete separation of the $\beta_i$ dependence is given inductively by
\begin{eqnarray}
M_k  = \frac{M _{k+1}}{\sin ^2 \beta _k} -\frac{\Psi^2 _{p-k+1}}{\cos ^2
\beta _k} + \frac{1}{\Phi _{\beta_k} \cos \beta _k \sin ^{2p-2k-1} \beta_k}
\frac{d}{d\beta _k} \left( \cos\beta _k \sin \beta _k \frac{d\Phi _{\beta
_k}}{d\beta _k} \right) \,,
\end{eqnarray}
where $k=1,...,p-1$, and we use the convention $M_p = -\Psi_1 ^2$.
These results agree with the previously known analysis in five dimensions \cite{frolov2}.

At this point we have complete separation of the Klein-Gordon
equation in the Kerr-(Anti) de Sitter black hole background in all
dimensions with two sets of possibly unequal rotation parameters
in the form given by (\ref{KGanz}) with the individual separation
functions given by the ordinary differential equations above. Note
that the separation of the Klein-Gordon equation in this geometry
is again due to the existence of the non-trivial Killing tensor.

\section*{Conclusions}
We studied the integrability properties of the Hamilton-Jacobi and
the massive Klein-Gordon equations in the Kerr-(Anti) de Sitter
black hole backgrounds in all dimensions. Complete separation of
both equations in Boyer-Lindquist coordinates is possible for the
case of two possibly unequal sets of rotation parameters. We
discuss the Killing vectors and reducible Killing tensors that
exist in the spacetime and also construct the nontrivial
irreducible Killing tensor which explicitly permits complete
separation. Thus we demonstrate the separability of the
Hamilton-Jacobi and the Klein-Gordon equations as a direct
consequence of the enhancement of symmetry. We also derive
first-order equations of motion for classical particles in these
backgrounds, and analyze the properties of some special
trajectories.

Further work in this direction could include the study of
higher-spin field equations in these backgrounds, which is of
great interest, particularly in the context of string theory.
Explicit numerical study of the equations of motion for specific
values of the black hole parameters could lead to interesting
results. The first order equations of motion presented here can
also readily be used in the detailed study of black hole
singularity structure in an AdS background geodesic probes and the
AdS/CFT correspondence.

\end{document}